\documentclass{article}

\PassOptionsToPackage{numbers, compress}{natbib}

\usepackage[final]{neurips_2020}

\usepackage[utf8]{inputenc} 
\usepackage[T1]{fontenc}    
\usepackage{hyperref}       
\usepackage{url}            
\usepackage{booktabs}       
\usepackage{amsfonts}       
\usepackage{nicefrac}       
\usepackage{microtype}      

\usepackage{amsmath}
\usepackage{amssymb}   
\usepackage{amsthm}   
\usepackage{bm}
\usepackage{float}
\usepackage{url}
\usepackage{graphicx}
\usepackage{enumitem}
\usepackage{hyperref}
\hypersetup{
  colorlinks   = true, 
  urlcolor     = blue, 
  linkcolor    = black, 
  citecolor   = black 
}

\newcommand{\be}{\begin{equation}}
\newcommand{\ee}{\end{equation}}

\title{Meta-modeling strategy for data-driven forecasting}

\author{%
  Dominic J.~Skinner\\
  Department of Mathematics\\ 
  Massachusetts Institute of Technology\\
  Cambridge, MA 02139-4307\\
  \texttt{dskinner@mit.edu}
  \And
  Romit Maulik \\
Argonne Leadership Computing Facility \\
Argonne National Laboratory \\
Lemont, IL 60439 \\
\texttt{rmaulik@anl.gov}
}

\begin{document}

\maketitle
\begin{abstract}
Accurately forecasting the weather is a key requirement for climate change mitigation. Data-driven methods offer the ability to make more accurate forecasts, but lack interpretability and can be expensive to train and deploy if models are not carefully developed. Here, we make use of two historical climate data sets and tools from machine learning, to accurately predict temperature fields. Furthermore, we are able to use low fidelity models that are cheap to train and evaluate, to selectively avoid expensive high fidelity function evaluations, as well as uncover seasonal variations in predictive power. This allows for an adaptive training strategy for computationally efficient geophysical emulation.
\end{abstract}

\vspace{-8pt}
\section{Introduction}
While numerical weather forecasting dates back over a century~\cite{WeatherPred}, climate change is expected to significantly alter the predictability of the atmosphere, increasing the error of weather forecasts~\cite{Scher2019}. Simultaneously, climate mitigation requires more accurate forecasts of weather events for power grid optimization~\cite{Auffhammer2016}, and extreme event prediction~\cite{SILLMANN2017}, from flooding~\cite{Kendon2014}, to heatwaves~\cite{Ford2018}. Even the largest climate simulations have $O(10\,\text{km})$ separation between grid points, meaning dynamics of smaller scales can not be explicitly resolved~\cite{mizuta2012climate,Satoh2008,Ohfuchi2005}, and often ad-hoc closure conditions are postulated to account for the unresolved dynamics. Machine learning has the potential to find data-driven closure conditions~\cite{rackauckas2020universal} and parameterize sub-grid scale modeling~\cite{Brenowitz2018,Gentine2018}. Convolutional neural networks can forecast weather~\cite{Chattopadhyay2020}, deep learning can predict extreme weather events~\cite{ExtremeWeatherPred}, and neural network architectures can be optimized automatically to enhance the quality of forecasts~\cite{GeophysicalEmulation,DeepHyper}.
However, these models can become expensive to train and deploy, especially when ensemble predictions are required~\cite{Gneiting2005}, and often their predictions lack interpretability. Here, we use a combination of proper orthogonal decomposition (POD), and long-short term memory (LSTM) reccurent neural networks to create forecasts for two real-world data sets. In addition to the LSTM network, which is our high-fidelity (HF) model, we introduce simpler or low-fidelity (LF) models, which do not have the same predictive power, but are faster to deploy. By studying the prediction differences of the LF models, we are able to selectively avoid HF function evaluations, as well as uncovering seasonal variations in prediction accuracy. With the proliferation of data driven methods for climate forecasting, the approach introduced here, could be used to reduce computational resources, which is of particular relevance for ensemble forecasting~\cite{Gneiting2005}, as well as to identify uncertainties in neural network predictions.

\section{Proper Orthogonal Decomposition}
To first reduce the dimensionality of the problem, we project spatially resolved fields, such as temperature, onto a set of principal modes which capture the salient features and then track the evolution of these modes in time. We use the technique of Proper Orthogonal Decomposition (POD) in order to find the dominant modes~\cite{ModalAnalysisRev, LSTMMaulik,MILANO2002}. In short, 
suppose we are given spatial snapshots of the data at various times, 
$\bm{\theta}_1, \dots, \bm{\theta}_T$. We take a truncated set of orthonormal basis vectors $\bm{v_1}, \dots, \bm{v}_M$, that approximates the spatial snapshots optimally with respect to the $L_2$ norm, i.e. minimizing $\sum_i^T || \bm{\theta_i} - \bm{\phi}_i||^2$
over $\bm{\phi}_i \in \text{Span}\{\bm{v}_1, \dots, \bm{v}_M\}$.
Defining the snapshot matrix, $\bm{S} = [ \bm{\theta}_1 | \cdots | \bm{\theta}_T ]$, the optimal basis is given by the $M$ eigenvectors of $\bm{SS}^T$, with largest eigenvalues~\cite{ModalAnalysisRev}, which is readily found numerically. See appendix~\ref{App:POD}, for further details.

\begin{figure}\centering
\includegraphics[width=\textwidth]{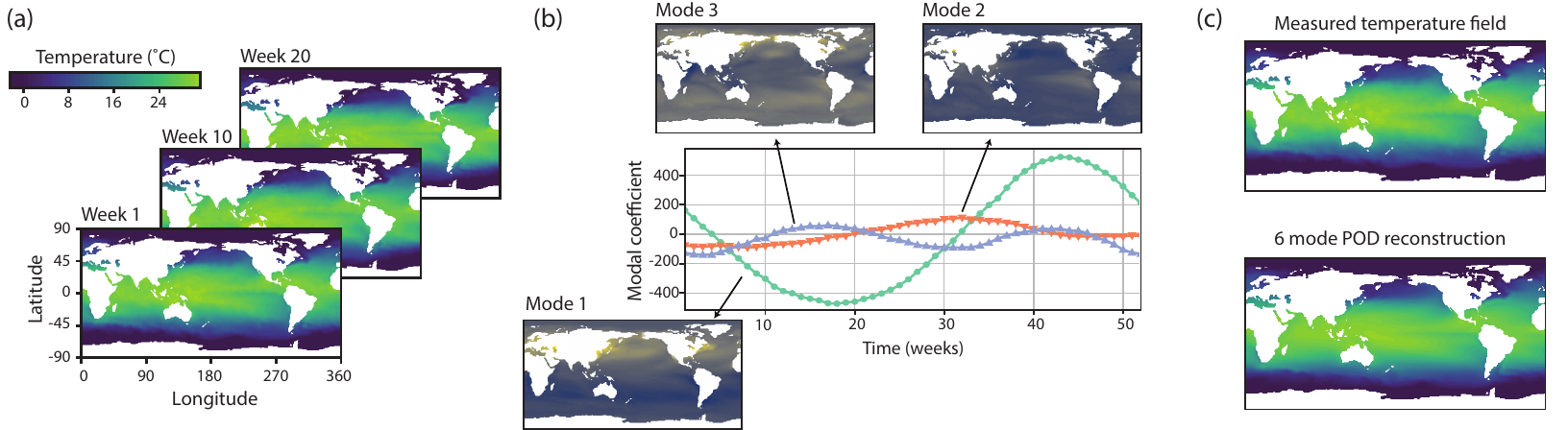}
\vspace*{-10pt}
\caption{\label{fig:ExplainFig} Proper orthogonal decomposition of sea surface temperature. (a) Historical climate data measuring ocean surface temperature weekly from 1981 to 2000. (b) Evolution of the first 3 modal coefficients, together with their respective basis elements. (c) Reconstructed temperature field from first the 6 modal coefficients against measured temperature field.}
\end{figure}

\section{NOAA sea surface temperature}
\begin{figure}
    \centering
    \includegraphics[width=0.9\textwidth]{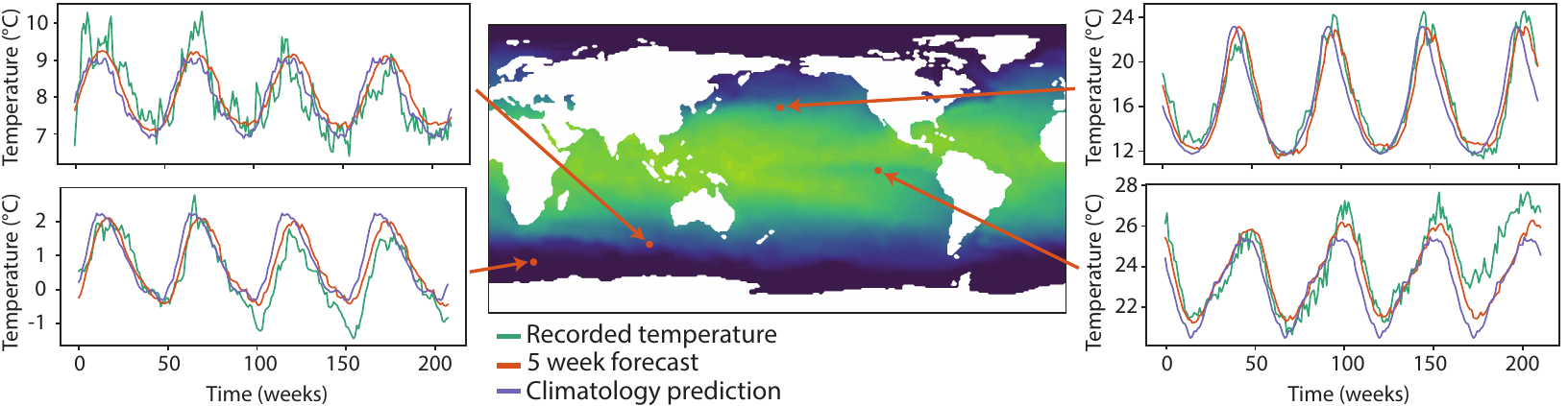}
    \caption{Sea surface temperature at 4 randomly chosen sensors (orange points) for the first 4 years of testing data (2002-2006), with POD-LSTM prediction and climatology baseline. After training, the LSTM model was applied non-autoregressively, making 5 weeks of predictions at a time. The spatial predictions were then reconstructed from the POD basis. Also shown is the climatology baseline.}
    \label{fig:SensorPred}
\end{figure}

Our first data set is the NOAA Optimum Interpolation SST V2 data set, containing the sea surface temperature weekly on a 1 degree grid across the period 1981-2018~\footnote{Available at \url{https://www.esrl.noaa.gov/psd/}}. The first 20 years were used as training data, the rest was reserved for testing. From this, we build data driven forecasts, predicting the next 5 weeks of temperature evolution from historical data. We project the system onto the first 6 POD modes, which is sufficient to approximate the temperature field (Fig.~\ref{fig:ExplainFig}), and examine the evolution of these modal coefficients (Fig.~\ref{fig:ExplainFig}a-b). 
From predictions of the modal coefficients, we can reconstruct the temperature field and test the temperature predictions at a sensor located at a specific coordinate, Fig.~\ref{fig:SensorPred}. The baseline to improve on is the climatology prediction; the historical average temperature at that specific location for the time of year.

\begin{figure}\centering
\includegraphics[width=\textwidth]{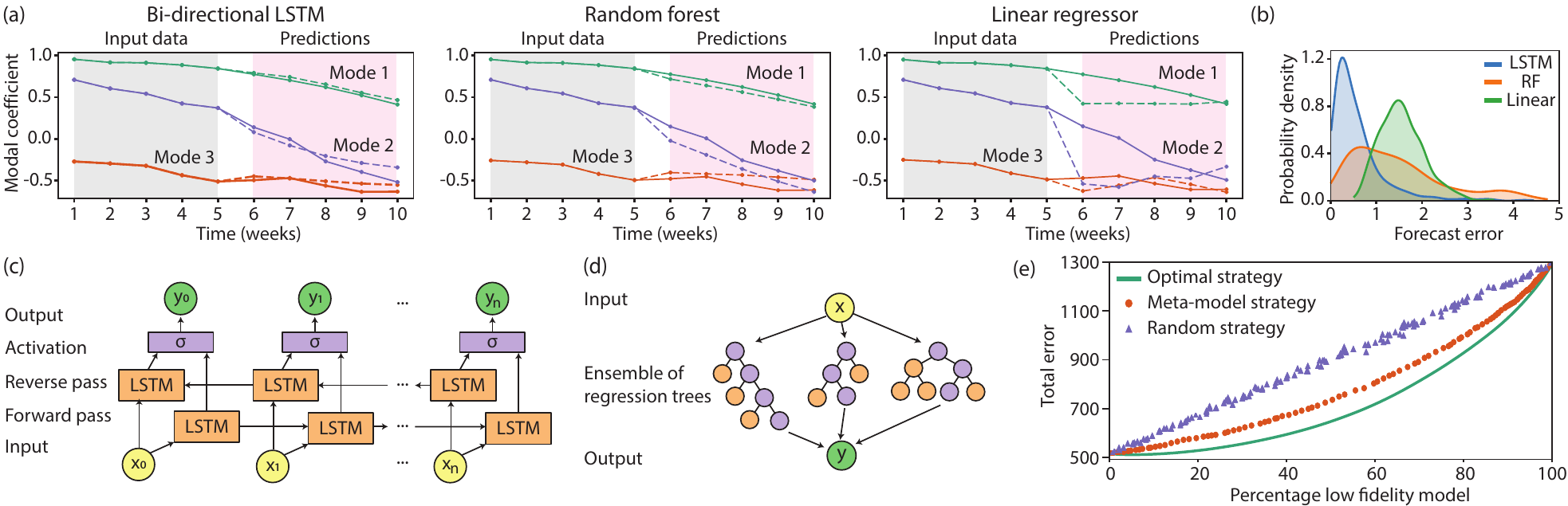}
\vspace*{-12pt}
\caption{\label{fig:ModelComparison} Comparison of high and low-fidelity models. (a) Input and predictions for a representative test example, true modal coefficients are solid lines, predictions are dashed. (b) Distributions of forecast errors for the testing data. (c) Architecture of the bi-directional LSTM, in our case $n=10$. The input information travels forwards through time, and then backwards again to allow for a more accurate global prediction. (d) Architecture of the random forest; the input is sent to several decision trees, and their result is aggregated at the end for a prediction. (e) Total error of adaptive meta-modeling strategy against the optimal and random strategy }
\end{figure}

For our purposes, we will use a bi-directional LSTM network~\cite{BiRNN} as our high-fidelity model, taking 5 weeks of historical data as input and making a 5 week forecast. This model is proto-typical for more complex machine learning models that could be deployed on data like this~\cite{GeophysicalEmulation}. In short, a bi-directional LSTM is a recurrent neural network, that instead of acting purely sequentially on the data, has an additional pass backwards in time, Fig.~\ref{fig:ModelComparison}c, in practice improving predictive power. In all cases, we take the first 20 years as training data, and use the remainder for testing. We compare the LSTM predictions against the recorded temperature and climatology baseline, Fig.~\ref{fig:SensorPred}, finding that while both the climatology and LSTM forecast captures the seasonal trends, the LSTM predictions have lower $L_2$ error and cosine norm closer to 1, as compared to the recorded temperature, than the climatology prediction for all sensors tested.

Our aim now is to investigate when and where the model breaks down, as well as identifying where expensive HF function evaluations can be avoided, and substituted for low-fidelity alternatives. To do so, we introduce two LF models. The first is linear regression, applied autoregressively to each mode individually, with an input window of 5 weeks, and making a single prediction. The best linear model is fitted to the training data for a particular mode, and then applied to the testing data. The second is a random forest regressor~\cite{RandomForest}, which takes in all 5 weeks of input, and trains a model to predict each modal coefficient across the forecast. This works by training 100 decision trees on the input sequence, and then taking an average of the results as the output, Fig.~\ref{fig:ModelComparison}d.

Unsurprisingly, the LF models perform worse on testing data than the HF one, Fig.~\ref{fig:ModelComparison}(a,b). However, the low-fidelity models are extremely cheap to train and evaluate, we therefore seek to make use of them whenever possible. Suppose now that we have a limited computational budget for evaluating a number of forecasts, and so are forced to use a mixture of HF and LF models. If we made the choice randomly
with probability $p$ for the HF model, we would expect the average error to be $p\times(\text{HF average error}) + (1-p) \times(\text{LF average error})$. Similarly, we expect the computational cost to be $p\times(\text{HF cost}) + (1-p) \times(\text{LF cost})$. We will use the two LF models to create a meta-modeling strategy that achieves lower error than the random strategy for the same computational cost. The strategy is as follows:
\begin{enumerate}[topsep=0pt,itemsep=-1ex,partopsep=1ex,parsep=1ex]
	\item Take input data and perform the random forest and linear forecasts.
	\item Calculate the difference between the predictions. 
	\item If this difference exceeds some threshold, evaluate the HF model.
		  If not, stick with the LF model.
\end{enumerate}
This simple strategy using two LF models significantly outperforms the random strategy, Fig.~\ref{fig:ModelComparison}e. Moreover, it performs similarly to the theoretical optimum strategy, defined as the best possible choice of HF and LF evaluations, given their values and the true value in advance. This optimum strategy is clearly not viable in practice, yet the meta-model strategy performs similarly.


\section{DayMet North America Daily Surface Temperature}
\begin{figure}\centering
\includegraphics[width=\textwidth]{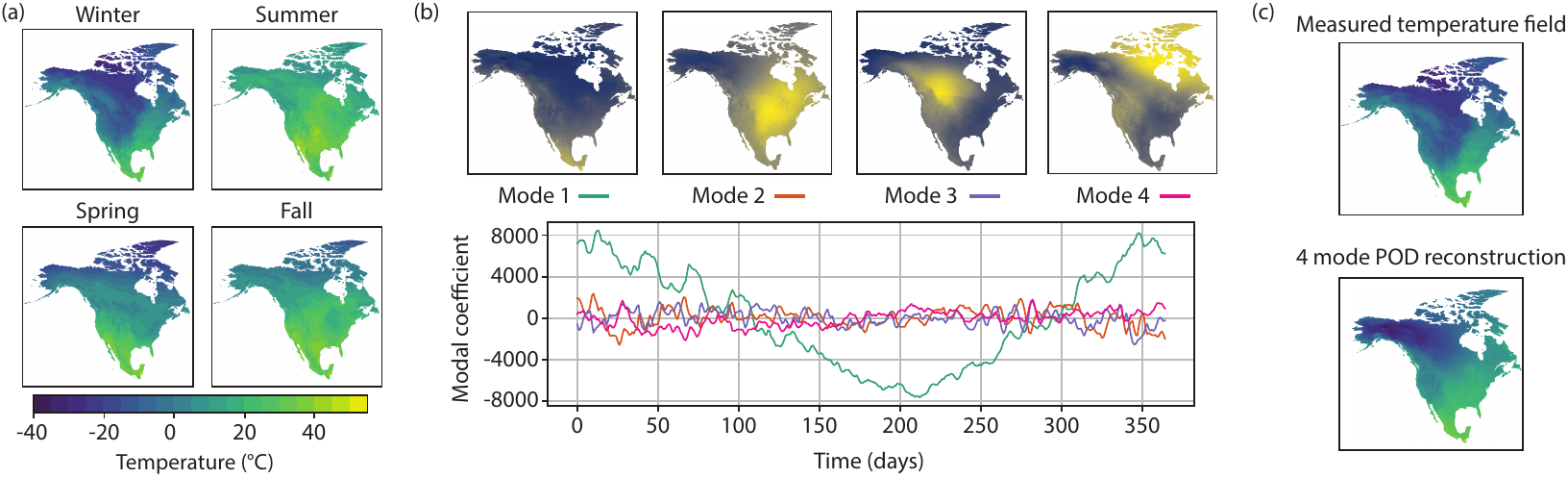}
\vspace*{-12pt}
\caption{\label{fig:ModalEvoNA} (a) Maximum daily temperature field across North America in 2011, representative seasonal snapshots shown. (b) Temporal evolution of the first four POD modes of the temperature field across one year, together with their representative basis elements. (c) Reconstruction of the temperature field from 4 modes against the true value.}
\end{figure}

We now validate the ideas we have introduced on an alternative data set. We consider the forecasting problem for the Daymet data set\footnote{available at \url{https://daymet.ornl.gov/}}, containing the maximum daily temperature field across North America for 2000-2015, Fig.~\ref{fig:ModalEvoNA}. The first 11 years are used as training data, the rest is reserved for testing. The temperature over land is more variable than the sea temperature, so here we forecast 7 days from 7 days of historical data, and take only the first 4 POD modes.

\begin{figure}\centering
\includegraphics[width=\textwidth]{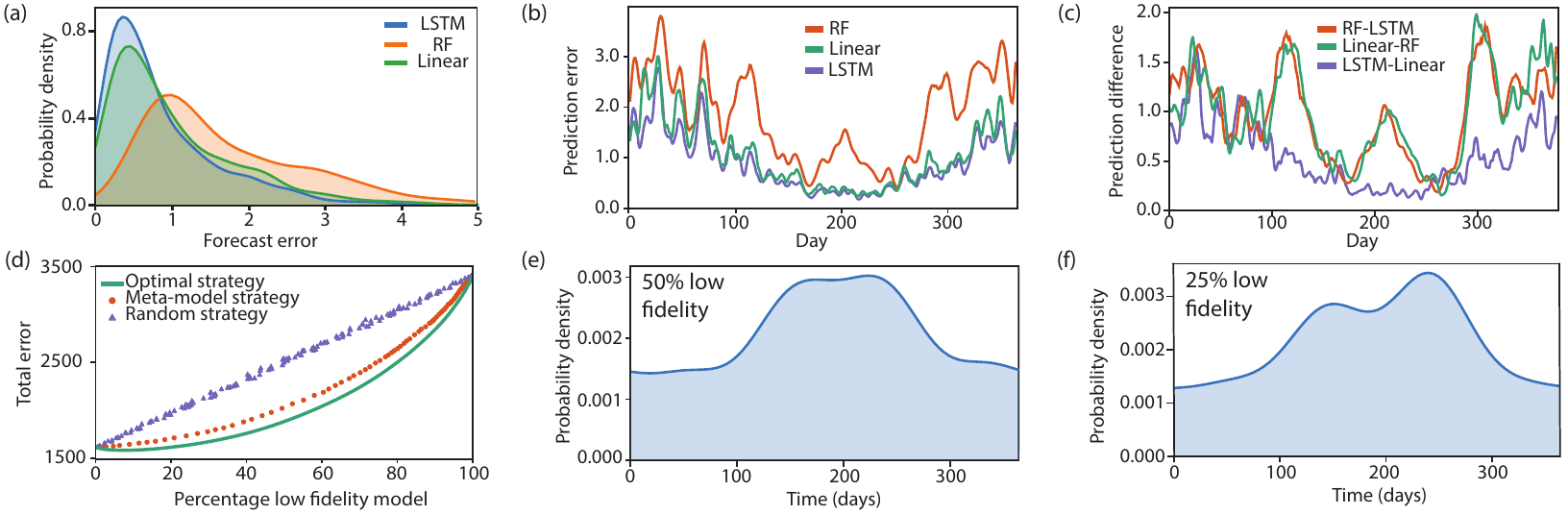}
\vspace*{-12pt}
\caption{\label{fig:SeasonalPred} (a) Forecasting error distribution for HF and LF models. (b) Seasonal prediction error, averaged across 5 years of testing data and over a window of 5 days. (c) Seasonal prediction difference between models. Model predictions were made, and the norm of their difference was calculated and averaged as in (b). Difference between LSTM and linear models recovers seasonal trend in error. (d) Total error of meta-modeling strategy against the theoretical optimum and random strategy. (e-f) Seasonal distribution for where the LF model is evaluated for $25\%$ and $50\%$ LF model evaluations. }
\end{figure}

The bi-directional LSTM has the lowest total prediction error, Fig.~\ref{fig:SeasonalPred}, and outperforms the climatology baseline for all sensors tested. The linear model performs almost as well, and both perform better than the random forest, Fig.~\ref{fig:SeasonalPred}a-b. Beyond overall error analysis, we aim to interrogate where  predictions break down. Averaging the prediction error across 5 years of testing data, we find that error varies with season, with the worst predictions in winter, Fig.~\ref{fig:SeasonalPred}c. Examining the difference between models, we also find this seasonal variation, meaning we need not know the error to 
deduce the regions of low predictive power, Fig.~\ref{fig:SeasonalPred}d. On this data, the meta-model approach is almost as good as the theoretical optimum strategy, Fig.~\ref{fig:SeasonalPred}d. Under this strategy, we see that in the summer when the LF models agree, a LF model is used for prediction, whereas in the winter the HF model is used, Fig.~\ref{fig:SeasonalPred}e-f.

\vspace{-1pt}
\section{Summary}
In this work, we considered the forecasting problem for two real-world geophysical data sets, the NOAA sea surface temperature and the NASA Daymet land surface temperature data for North America. We confirmed that the combination of dimensionality reduction through proper orthogonal decomposition and recurrent neural network based predictions outperforms a climatology baseline. While the neural network prediction outperforms simple random forest and linear predictions, we were able to extract information from these LF models nevertheless. Specifically, by using the two LF models, we were able to devise a simple rule for deciding whether to use a HF or LF model, which allowed a near optimal decision strategy. When used as a strategy for complicated architectures, this could allow substantial computational savings, especially when ensemble predictions are required. We were further able to use the LF models to understand seasonal variation in predictive power, finding that our models predictive power is highest in the summer for land surface temperature, a valuable insight for climate modeling. In future, it will be of interest to apply this to more advanced network architectures. 

\section*{Acknowledgments}

This material is based upon work supported by the U.S. Department of Energy (DOE), Office of Science, Office of Advanced Scientific Computing Research, under Contract DE-AC02-06CH11357. This research was funded in part and used resources of the Argonne Leadership Computing Facility, which is a DOE Office of Science User Facility supported under Contract DE-AC02-06CH11357. DS acknowledges support from the NSF-MSGI fellowship. This paper describes objective technical results and analysis. Any subjective views or opinions that might be expressed in the paper do not necessarily represent the views of the U.S. DOE or the United States Government.

\appendix
\section{Proper Orthogonal Decomposition}\label{App:POD}
Proper orthogonal decomposition (POD) provides a systematic method to project dynamics of a high dimensional system onto a lower dimensional subspace. We suppose that a single snapshot of the full system is a vector in $\mathbb{R}^N$, where $N$ could be the number of grid points at which the field is resolved. Observing the system across a number of time points gives us the snapshots $\bm{\theta}_1, \dots, \bm{\theta}_T$, with mean subtracted by convention. The aim of POD, is to find a small set of orthonormal basis vectors $\bm{v_1}, \dots, \bm{v}_M$, with $M \ll N$, which approximates the spatial snapshots,
\be
\bm{\theta}_t \approx \sum_{j=1}^M a_j(t)\bm{v}_j, \quad t=1,\dots,T,
\ee
and so allows us to approximate the evolution of the full $N$ dimensional system, by considering only the evolution of the $M$ coefficients $a_j(t)$. POD chooses the basis, $\bm{v}_j$, to minimize the residual with respect to the $L_2$ norm, 
\be
R = \sum_{t=1}^{T} || \bm{\theta}_t - \sum_{j=1}^M a_j(t) \bm{v}_j ||^2.
\ee
Defining the snapshot matrix, $\bm{S} = [ \bm{\theta}_1 | \cdots | \bm{\theta}_T ]$, the optimal basis is given by the $M$ eigenvectors of $\bm{SS}^T$, with largest eigenvalues, after which, the coefficients are found by orthogonal projection, $a_j(t) = \langle \bm{\theta}_t,\bm{v}_j\rangle$~\cite{ModalAnalysisRev}. 

For both of our data sets, we take only the training data snapshots, say $\bm{D}_1, \dots,\bm{D}_T$, from which we calculate the mean $\bar{\bm{D}} = (1/T) \sum_t \bm{D}_t$, hence defining the mean subtracted snapshots $\bm{\theta}_t = \bm{D}_t - \bar{\bm{D}}$. We then create the snapshot matrix, $\bm{S}$, and find numerically the $M$ eigenvectors of $\bm{S}\bm{S}^T$ with largest eigenvalues. From this, we train models, $\mathcal{N}$, to forecast the coefficients
\be
\bm{a}(t+1) \approx \hat{\bm{a}}(t+1) = \mathcal{N} ( \bm{a}(t), \bm{a}(t-1),\dots).
\ee
making predictions of future coefficients given previous ones. 

To test the predictions on unseen data, $\bm{E}_1,\dots,\bm{E}_k$, we take the mean $\bar{\bm{D}}$, and vectors $\bm{v}_j$ calculated from the training data to get that 
\be
a_j(t) = \langle \bm{E}_t - \bar{\bm{D}}, \bm{v}_j \rangle, \quad j = 1,\dots,M,
\ee
which will be used by the model $\mathcal{N}$ to make a prediction for future coefficients. The prediction for the coefficients $\hat{\bm{a}}$, can be converted into predictions in the physical space by taking $\bar{\bm{D}} + \sum_j \hat{a}_j \bm{v}_j$. This procedure only makes use of testing data to pass into the model, not to train the model in any way. Crucially, to make a forecast of $\bm{E}_{t+1}$, 
only previous measurements $\bm{E}_{t},\bm{E}_{t-1}, \dots$ are needed.

\end{document}